\documentclass[a4paper,review,number]{elsarticle}
\usepackage{graphicx}
\usepackage{amssymb}
\usepackage[usenames]{color}

\usepackage[top=1.5in, bottom=1.25in, left=1in, right=1in]{geometry}

\begin{document}

\title{Short-Lived Electron Transfer in Donor-Bridge-Acceptor Systems}

\author[rvt]{D.~Psiachos\corref{cor1}}
\ead{dpsiachos@gmail.com}
\cortext[cor1]{Corresponding author}
\address[rvt]{Department of Physics, University of Crete, Heraklion 71003, Greece}


\begin{abstract}
We investigate time-dependent electron transfer (ET) in benchmark donor-bridge-acceptor systems. For
the small bridge sizes studied, we obtain results far different from the perturbation theory which underlies
scattering-based approaches, notably a lack of destructive interference in the ET for certain
arrangements of bridge molecules. We also calculate wavepacket transmission in the non-steady-state regime, finding a featureless 
spectrum, while for the current we find two types of transmission: sequential and direct, where in the
latter, the current transmission \emph{increases} as a function of the energy of the transferred electron,
a regime inaccessible by conventional scattering theory.
\end{abstract}

\maketitle

\section{Introduction}
Electron transfer (ET) in donor-bridge-acceptor (D-B-A) systems composed of organic molecules or molecular chains has been studied actively
for the past few decades due to the potential applications to molecular electronics, DNA, etc. Expressions for
estimating ET rates,
closely-related to those derived for the superexchange mechanism~\cite{Anderson},
specifically applied to ET by the McConnell expression~\cite{McConnell}, have found extensive
application especially for biological
systems~\cite{rps,Marcus,Ogrodnik} and they continue to be widely 
used for such systems (\textit{e.g.} photosynthesis, proteins, DNA)~\cite{Skourtis, Kirk, Ratner, Beratan, Xiao,Zarea}. 

The Landauer formula states that the electronic conductance through a molecule connected to two leads 
is proportional to the transmission, a quantity which depends on the scattering processes
involved. Some of the methods used to describe the scattering include Green's function methods,
transfer matrix approaches~\cite{esqc}, or the use of
Lippmann-Schwinger scattering operators for mapping the non-equilibrium system onto an equilibrium
one~\cite{han}. All of these
methods are time-independent, thus enabling the use of the Landauer or, in the case of several leads, the
Landauer-B\"{u}ttiker formula, for evaluating the current. 

Most studies of electron transmission through bridges,~\textit{e.g.} Ref.~\cite{frederiksen},
are in fact based on non-equilibrium Green's function methods rather than explicitly 
time-dependent calculations. Recently however,
time-dependent approaches are increasingly being applied. For instance the wave-packet dynamics approach, wherein
a spatially-Gaussian wavepacket with the Fermi energy of a scanning-tunnelling microscope tip bulk is launched towards
the tip apex and tunnels into the sample~\cite{WPdynamics}. Time-dependent
approaches for studying the transfer of a wavepacket through
organic molecules attached to electrodes~\cite{Bishop} using Green's function formalism have also been used
in combination with perturbation theory to study the transfer in the presence of dissipation~\cite{Renaud}. In such
wavepacket approaches though, the transmission function is simply a Fourier-transform of the system (tip-sample or 
molecule+electrodes). This is to be contrasted with an
explicit inclusion of the wavepacket as part of the system~\cite{Psiachos2014} where the eigenvalue
spectrum can be modified by an incident electron off-resonance, leading to effects such as D-A tunnelling. Time-dependent forms of the transfer-matrix element owing
to lattice effects were derived in Ref.~\cite{Skourtis2} while a detailed comparison of the ET mechanisms 
identified the ``through-bridge" contribution (`sequential' here) as a competing mechanism to that of the superexchange contribution~\cite{Skourtis3} (`direct' ET here). 

In
all af the scattering methods described above, the standard implementation is that the scattered electron 
does not modify the existing
energy level structure. Nonetheless, a vast body of literature exists where the influence of the transferred electron 
has been taken into account for calculating the Green's function. In most cases, one desires only the projection onto a two-state
approximation for D-A ET as a representation of the `direct' ET occurring, even though, as shown later, this quantity does not
in general capture all of the time-dependent phenomena. In a recent study~\cite{Nishioka}, perturbation theory was compared
with a self-consistent two-state construction based on the L\"owdin projection~\cite{Lowdin}. Further back, an analytical, exact,
expression for ET was derived for general bridge systems~\cite{Evenson}, similar to the work presented here for specific
models. Numerically, the Green's function for large systems has been calculated using non-perturbative techniques~\cite{Kakitani1}
and a procedure for an exact solution has been proposed~\cite{Onuchic1991}. Also, the approximations underlying the McConnell transfer-rate expression, 
have been assessed and various degrees of improvements proposed~\cite{Reimers}. In line with the latter is the work by Stuchebrukhov 
(\textit{e.g.} Ref.~\cite{Stuchebrukhov})
where the focus is on T-matrix expansions in order to achieve more accurate two-state pictures. However, the iterative techniques
aimed at improving the accuracy of the Green's function are confined to have certain regimes of validity,
and fail near the resonant-transfer limit~\cite{SkourtisOnuchic, SkourtisBeratan,Priyadarshy}. In addition, they
are not general enough, in assuming equal donor and acceptor energies.

While an accurate calculation of the exact energy-level structure leads to improved 
rate expressions, at least for direct ET, for conductance
calculations the scattering approach cannot be relied upon in cases where the ET is a transient phenomenon.
In a series of papers~\cite{AmosBurrows1, AmosBurrows2, AmosBurrows3} the current through 
small bridge molecules has been calculated time-dependently, in the presence of an electric field. Most 
notably, agreement was found with scattering theory in the steady-state regime~\cite{AmosBurrows3}, which 
nevertheless may be inapplicable to the small time scales of
interest in ET in many cases. Recently, the Multi-configuration time-dependent Hartree method has been applied to
study coherence in ultrafast electron transfer, including electronic-vibrational coupling~\cite{Thoss2015} while
this method has been compared with a time-dependent Green's function method within a reduced density matrix
formalism in Ref.~\cite{ThossPRB} to span a wide range of time scales. 

In this work, we correct the superexchange contribution at first instance, by
deriving exact expressions for D-A ET for simple benchmark systems. We 
then elaborate using full time-dependent calculations of ET through the system attached to electrodes on
the meaningfulness of the superexchange (what we refer to as ``direct") versus sequential ET as
the energy of the transferred electron is varied. In our
work, we do not aim to achieve the ``steady-state" regime but we compare our results with the steady-state
approaches, finding large differences. We further predict tunnelling-mediated
increases in current transmission, which are completely inaccessible by the scattering model.

\section{Theoretical Methods}
\begin{figure}[htb]
\includegraphics[width=14cm]{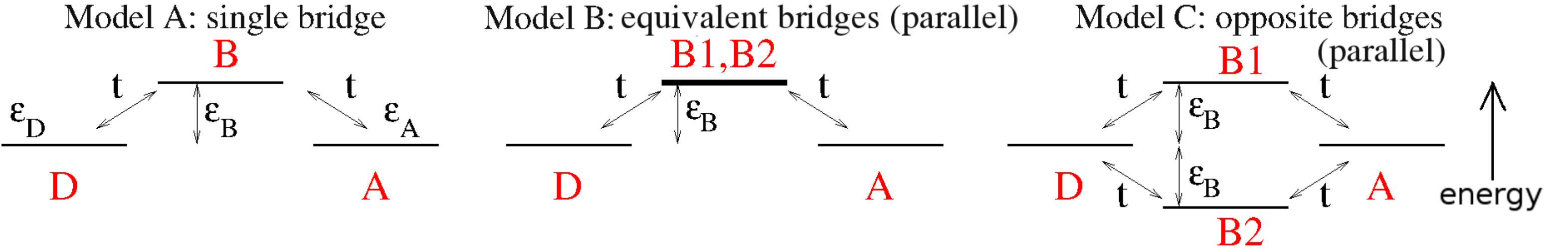}
\caption{The three model Donor-Bridge-Acceptor (D-B-A) systems under study. The
donor energy $\epsilon_D$ is depicted as equal to the acceptor energy $\epsilon_A$ for clarity.}
\label{dba}
\end{figure}

We aim at an exact solution for ET in the D-B-A systems shown in Fig.~\ref{dba}, in the spirit of Ref.~\cite{Evenson}
who considered exact approaches for deriving two-state effective Hamiltonians for general D-B-A systems. They consist
of a single bridge (Model A), two equivalent though non-interacting bridges (parallel configuration, Model B),
and two oppositely-configured bridges (Model C). The bridge energy is offset in general by $\left|\epsilon_B\right|$
from the donor and acceptor energy levels, which have been made equal for simplicity. Movement of electrons
between the bridges and the donor or acceptor levels is accomplished by a hopping energy $t$. 
In a later section (Sec.~\ref{sec:CurrentDensity}) we calculate the current transmission through the D-B-A 
system when attached to electrodes. 

The Hamiltonian in
second-quantized form describing these systems is given by
\begin{equation}
H=\sum_{i,\sigma}\epsilon_i c^\dagger_{i,\sigma}c_{i,\sigma}-t\sum_{\langle i,j\rangle,\sigma}\left(c^\dagger_{i,\sigma}c_{j,\sigma}+c^\dagger_{j,\sigma}c_{i,\sigma}\right)
\end{equation}
where $\langle i,j\rangle$ denotes that the hopping with energy $t$ is restricted to nearest neighbours, 
while the $\epsilon_i$ term describes the on-site energy of 
the electron. 

In the site basis $(|D\rangle,|A\rangle,|B\rangle)$ (the order 
between B and A having been reversed to write it in block form), $H$ for
the system depicted in Model A of Fig.~\ref{dba}
is 
\begin{equation}
H=\left(
\begin{array}{cc|c}
 \epsilon_D  & 0 & -t \\
 0 & \epsilon_A  & -t \\
\hline
 -t & -t & E_{B}
\end{array}
\right)
\label{fullH}
\end{equation}
where $E_B$ is the bridge energy.
Explicitly, in terms of the electron occupation
on the three sites (D-B-A) of Model A, the basis is $|D\rangle\equiv|\uparrow,\,\cdot,\,\cdot\rangle,$
$|B\rangle\equiv|\cdot,\,\uparrow,\,\cdot\rangle,$
$|A\rangle\equiv|\cdot,\,\cdot,\,\uparrow\rangle.$

By itself, a diagonalization of Eq.~\ref{fullH} gives no information in general on the direct transfer from D to A. In
contrast, for superexchange~\cite{Psiachos2015}, a direct diagonalization
yields the lowest singlet and triplet eigenstates, which are used to determine the superexchange interaction $J$. To 
study ET using Eq.~\ref{fullH},
an electron needs to be set up at time $t=0$ at site $D$, and its time evolution to $A$
studied using the time-dependent Schr\"{o}dinger equation (TDSE). 
Not in the least because the number of $B$ sites could be extremely large,
it is frequently preferred to extract approximate parameters for describing ET
using effective Hamiltonians.
Thus, we may use one of the well-known procedures for constructing an effective Hamiltonian
between D and A and examine the off-diagonal, transfer-matrix element. In simple
rate expressions \textit{e.g.} Refs.\cite{McConnell,Marcus} the rate is proportional 
to the squared magnitude of the transfer-matrix element. In more sophisticated models the electron-nuclei
interactions are treated in greater detail~\cite{Skourtis2,Kakitani2,BlumbergerChemRev}.

\subsection{Effective Hamiltonian}
The blocks of the Hamiltonian Eq.~\ref{fullH} are denoted by the following short-hand notation:
\begin{equation}
H=\left(\begin{array}{c|c}
\underline{\underline{H_{00}}} & \underline{\underline{T_{01}}}\\
\hline
\underline{\underline{T_{10}}} & \underline{\underline{H_{11}}}
\end{array}\right).
\label{block}
\end{equation}

The purpose of defining an effective Hamiltonian is to describe the full Hamiltonian
in terms of a subset of the full basis, in this case the $\underline{\underline{H_{00}}}$ block. An exact projection may be obtained
from a L\"{o}wdin approach~\cite{Lowdin}. The solution, in terms of the notation of Eq.~\ref{block},
is given by
\begin{equation}
H_{\mathrm{eff}}(E)=\underline{\underline{H_{00}}}+\underline{\underline{T_{01}}}\left(E\underline{\underline{I}}-\underline{\underline{H_{11}}}\right)^{-1}\underline{\underline{T_{10}}}.
\label{Heff}
\end{equation}

Explicitly, and after making the system symmetric and
redefining the energy zero for simplicity in comparing
with literature results; $\epsilon_D=\epsilon_A\equiv 0$, Eq.~\ref{Heff} is given by
\begin{equation}
H_{\mathrm{eff}}(E)=\left(\begin{array}{c c}
\frac{t^2}{E-\epsilon_B} & \frac{t^2}{E-\epsilon_B}\\
\frac{t^2}{E-\epsilon_B} & \frac{t^2}{E-\epsilon_B}
\end{array}\right),
\label{Heff1}
\end{equation}

while our eigenvalue problem is now
\begin{equation}
H_{\mathrm{eff}}(E)X=EX,\;\;\;X =\left(c_1\left|D\rangle\right.,c_2\left|A\rangle\right.\right),
\label{eigHeff}
\end{equation}
that is the eigenvectors are a linear combination of D and A states.

The quantity $E$ is the ET energy and it can be determined exactly.
It can lead to the exact eigenvalue for \emph{one} element at a time in $X$ if it is set to the corresponding
eigenvalue of Eq.~\ref{eigHeff}. However, most of the time it is set to the unperturbed eigenvalue,
in this case $\epsilon_D$ or $\epsilon_A$, as per Rayleigh-Schr\"{o}dinger perturbation theory~\cite{Lindgren,BWbook},
with the aim of using one effective Hamiltonian for describing all solutions within an energetically-degenerate subspace.
In this work, we will show that this leads to qualitatively incorrect behaviour, while Model C is completely
untreatable by perturbation theory.

By solving Eq.~\ref{eigHeff} three distinct roots, labelled $\lambda_1,\lambda_2,\lambda_3,$ are obtained,
exactly equal to the eigenvalues of the full H, Eq.~\ref{fullH}, for all parameter values. 

\begin{figure}[htb]
\includegraphics[width=6cm]{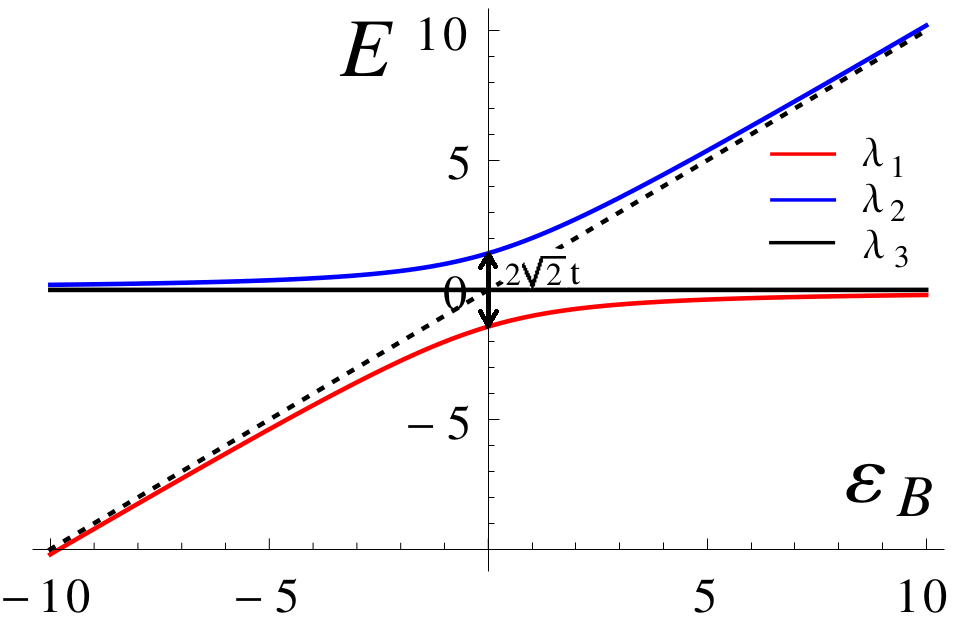}
\caption{The three energy eigenvalues of Eq.~\ref{Heff1} as the relative bridge energy is varied for
 $\epsilon_D=\epsilon_A.$ The dashed line is an asymptote: $E=\epsilon_B,$ shown
as a guide to the eye. Scales in units of $t$.}
\label{root123}
\end{figure}

Fig.~\ref{root123} shows the three roots as the bridge energy relative 
to the ends (which are set equal: $\epsilon_D=\epsilon_A$)
is varied. 

\section{D-A Effective Transfer}
Here the results for the off-diagonal term of Eq.~\ref{Heff1}, hereafter
denoted by $T_{DA}$,
when $E$ is replaced by the three exact eigenvalues (Fig.~\ref{root123}) are shown, again
assuming the symmetric case $\epsilon_D=\epsilon_A\equiv0,$ although this method works for general parameters. 

The elements in Eq.~\ref{Heff1} are not in general second-order in $t$; this is true only
for $t/\left|\epsilon_B\right|\ll1$ for the symmetric case studied here.
\begin{figure}[htb]
\includegraphics[width=14cm]{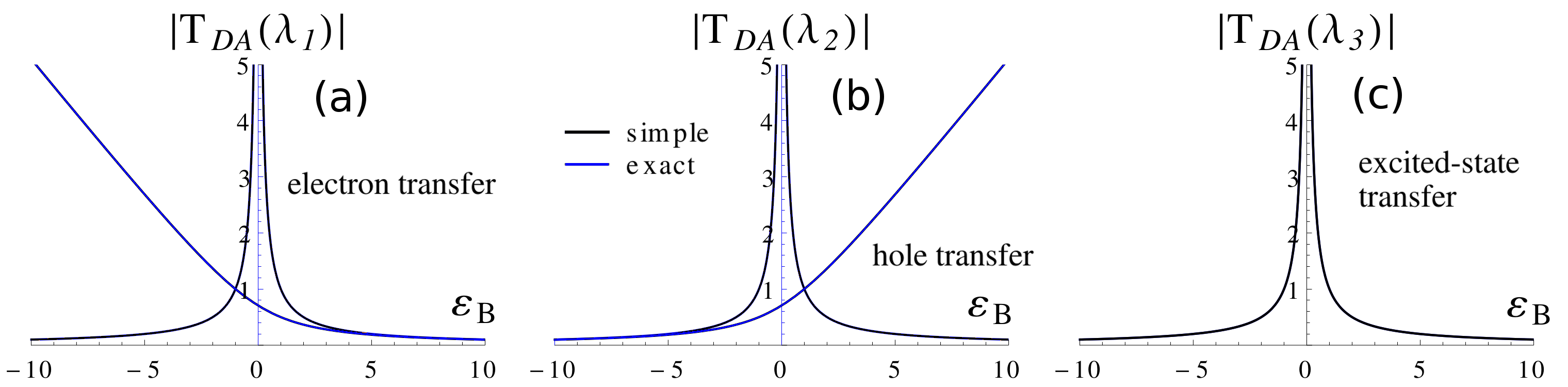}
\caption{The effective transfer $\left|T_{DA}\right|$ according to the each of the eigenvalues
(see Fig.~\ref{root123}) in two forms: the `simple' and the exact form (see text). For $\lambda_3$ the
two forms coincide. Scales in units of $t.$}
\label{tda123}
\end{figure}

In Fig.~\ref{tda123} there are two forms of the effective transfer shown: the `simple' form from using
the unperturbed D/A energies in Eq.~\ref{Heff1} and the exact form. Fig.~\ref{tda123}a depicts ET 
as it was derived using the lowest
eigenvalue. Fig.~\ref{tda123}b corresponds to hole transfer, which dominates
when $\epsilon_B<0$. For $\epsilon_B\ne 0$ we have contributions
from both; depending on their relative signs, either electron or hole transfer dominates,
while at resonance $\epsilon_B=0,$ the two contributions are equal. 

Indeed, a full time-dependent analysis of D-A transfer from solving the TDSE shows that the
transfer of a wavepacket originating on D contains three time-scales, or Rabi
frequencies. The occupation probability at A may be written as
\begin{equation}
|\psi_A(t)|^2=\sum_{i=1}^3 A_i\cos{2\omega_i t}
\label{rabi3}
\end{equation}
where the $\omega_i$ are equal to $T_{DA}(\lambda_1)$ for ET, to $T_{DA}(\lambda_2)$ for 
hole transfer, or to the sum of the two. None of the Rabi frequencies $\omega_i$ however is 
equal to $T_{DA}(\lambda_3)$, which 
corresponds to excited-state transfer of a wavepacket. 

As well, the functions $T_{DA}(\lambda_1)$
and $T_{DA}(\lambda_2)$ are finite-valued
at $\epsilon_B=0$ unlike the expression $T_{DA}(\lambda_3)$ (Fig.~\ref{tda123}).
In Fig.~\ref{tda123} the exact and `simple' expressions coincide, although $\lambda_3$ is not the lowest
eigenvalue. $\lambda_3$ is degenerate with one of the other two roots for large $\left|\epsilon_B\right|$
only and therefore the results of $T_{DA}$ derived using this root are invalid at smaller
$\left|\epsilon_B\right|$. Specifically, when $\epsilon_B=0$ the exact results for $T_{DA}$
are $t/\sqrt{2}$, $-t/\sqrt{2}$, and $\infty$ for the three roots $\lambda_1,$ $\lambda_2,$ and $\lambda_3$
respectively.

These results generalize for a bridge composed of many sequential sites.

\subsection{Independent paths}
Without loss of generality, we consider two independent paths from 
D-A, seen for example in Fig.~\ref{dba} Models B,C. We again treat the
donor and acceptor energy levels as equal for simplicity. However,
the two bridges may be non-equivalent. We take the extreme cases where
the two bridges have the same energy, or where one energy is the negative
of the other. Such situations would arise from considering bonding
or anti-bonding configurations of $\pi$
orbitals or more complex molecules~\cite{Vazquez}. Such general forms of bridge structures and the consideration of independent
pathways are, for example, applicable to proteins and
to photosynthesis molecules~\cite{Ratner, Beratan, Zarea}.

The Hamiltonian is written, again assuming $\epsilon_D=\epsilon_A\equiv 0$ as
\begin{footnotesize}{\begin{equation}
H=\left(
\begin{array}{cc|cc}
 0  & 0 & -t&-t \\
 0 & 0  & -t&-t \\
\hline
 -t & -t & \epsilon_{B1}&0\\
 -t & -t & 0&\epsilon_{B2}
\end{array}
\right) \begin{array}{c} \left| \uparrow,\,\cdot,\,\cdot,\,\cdot\rangle\right.\equiv|D\rangle\\ \left| \cdot,\,\cdot,\,\cdot,\,\uparrow\rangle\right.\equiv|A\rangle\\\left| \cdot,\,\uparrow,\,\cdot,\,\cdot\rangle\right. \equiv|B1\rangle\\\left| \cdot,\,\cdot,\,\uparrow,\,\cdot\rangle\right. \equiv|B2\rangle\end{array}
\end{equation}}
while the result of the projection onto the model space $\left( \left|D\rangle,\right.\left|A\rangle\right.\right)$ is
\begin{equation}
H_{\mathrm{eff}}(E)=\left(\begin{array}{c c}
\frac{2t^2}{E-\epsilon_B} & \frac{2t^2}{E-\epsilon_B}\\
\frac{2t^2}{E-\epsilon_B} & \frac{2t^2}{E-\epsilon_B}
\end{array}\right)\begin{array}{c}|D\rangle\\|A\rangle\end{array}
\label{Eb1b2}
\end{equation}
\end{footnotesize}
for equivalent bridge levels $\epsilon_{B1}=\epsilon_{B2}\equiv\epsilon_B$
and
\begin{equation}
H_{\mathrm{eff}}(E)=\left(
\begin{array}{cc}
 \frac{2 {E} t^2}{{E}^2-\epsilon_B^2} & \frac{2 {E} t^2}{{E}^2-\epsilon_B^2} \\
 \frac{2 {E} t^2}{{E}^2-\epsilon_B^2} & \frac{2 {E} t^2}{{E}^2-\epsilon_B^2}  \\
\end{array}\right)\begin{array}{c}|D\rangle\\|A\rangle\end{array}
\label{Eul}
\end{equation}
for opposite bridge levels, $\epsilon_{B1}=-\epsilon_{B2}\equiv\epsilon_B$ (Fig.~\ref{dba}).

\begin{figure}[htb]
\includegraphics[width=12cm]{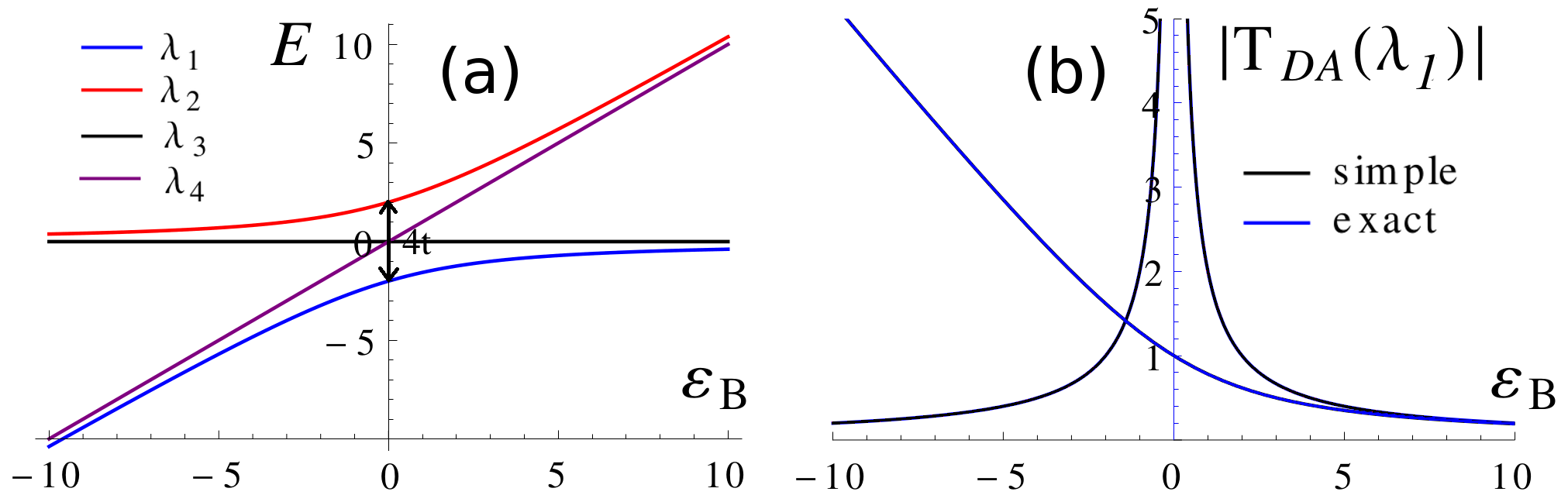}
\caption{(a) The eigenvalues of the equivalent-bridge situation (Eq.~\ref{Eb1b2})
as a function of the bridge energy. (b) The effective transfer $\left|T_{DA}\right|$ according to the lowest eigenvalue of the
equivalent-bridge case in the `simple' as well as the exact form (see text). $\epsilon_D=\epsilon_A=0$ has been assumed. The energy
scales are in terms of $t.$}
\label{tba1all}
\end{figure}

In the first case, of equivalent bridges (Model B), the results are quite similar to the single-bridge
case. The eigenvalues of Eq.~\ref{Eb1b2} and the corresponding transfer element for the lowest
root are shown in Figs.~\ref{tba1all}a and b where again the `simple' expression obtained using the 
unperturbed D/A energies in Eq.~\ref{Eb1b2} is included for comparison.

As in the single-bridge case, the transfer element only 
asymptotically approaches the `simple' expression, where
the unknown energy $E$ in Eq.~\ref{Eb1b2} was set equal to the unperturbed
energy $\epsilon_D=\epsilon_A=0.$ In the `simple' form, the transfer
element for the two independent bridges is twice that
of the single bridge: $\left|\frac{2t^2}{\epsilon_B}\right|,$ but this is not the
case when the exact expressions for $E$ are used.

\begin{figure}[htb]
\includegraphics[width=12cm]{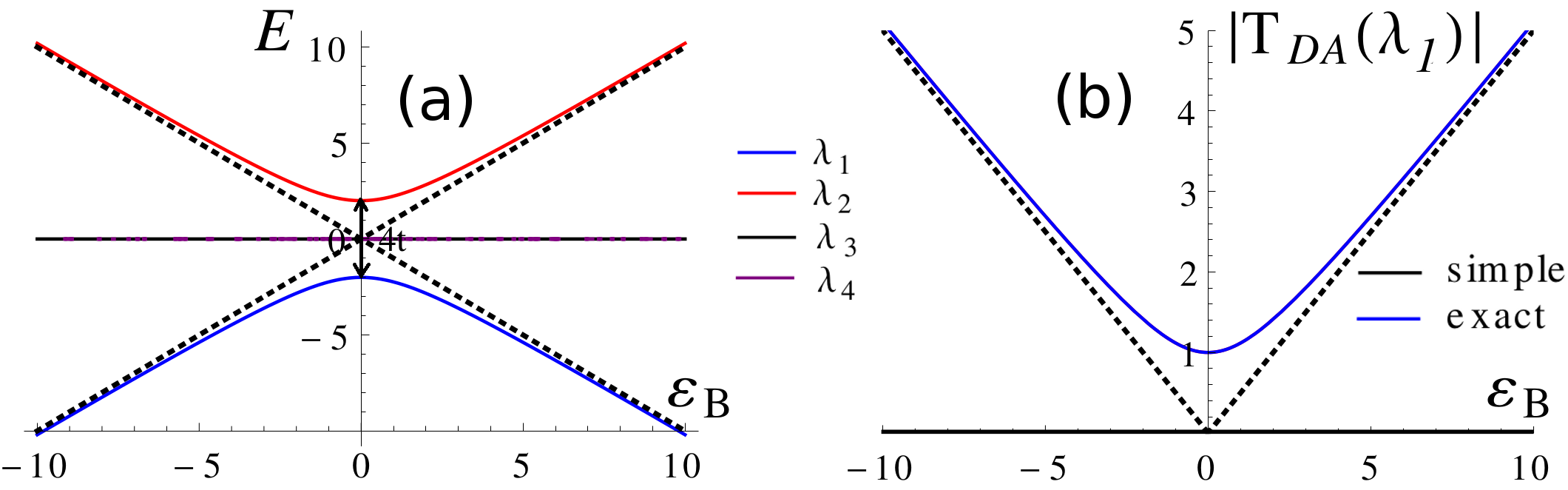}
\caption{The eigenvalues and corresponding transfer parameters for
the opposite-bridge case where $\epsilon_D=\epsilon_A=0$ has been assumed. The 
eigenvalues $\lambda_3$ and $\lambda_4$ in (a) are
always zero. Similarly, the `simple' expression in (b) is zero. Dashed lines
correspond to the asymptotes $E=\pm\epsilon_B$ and 
$\left|T_{DA}\right|=\pm\frac{1}{2}\epsilon_B$ which are guides
to the eye. Scales in terms of $t.$}
\label{2boppe}
\end{figure}

In the case of bridge energy levels having opposite signs (Model C), we get
the eigenvalues and transfer elements of Fig.~\ref{2boppe}.
Clearly, there is no asymptotic approach of the upper/lower eigenvalues
to any of the other roots. The use of the unperturbed energy results
in destructive interference, or $T_{DA}=0,$ as can be seen by Eq.~\ref{Eul} 
but this value is never accessed, except perhaps in an excited-state form of transfer.

\section{Transmission Characteristics}
The significance of the transfer matrix element is that it 
approximates ET by a single
Rabi frequency, which is exactly the dominant
Rabi frequency of the time-dependent 
solution of the full H. However, it tells us nothing about the
amplitude of the transfer. In Fig.~\ref{2boppe}
the transfer element gets \emph{larger} as the bridge energies
move off resonance, meaning
that in the simplest of models~\cite{McConnell,Marcus} the transfer rate is increased. However, the actual
amplitude of this dominant contribution to the ET (the coefficient $A_i$ of 
this term in Eq.~\ref{rabi3}) is vastly reduced, with the consequence that the rate of ET is reduced.

This is why quantities such as,
the wavepacket transmission and the current transmission for the D/A system attached to electrodes are more meaningful.
Notably, the current combines both qualities: rate and amplitude while the conductance
measured in experiments is proportional to transmission.

\begin{figure}[htb]
\includegraphics[width=10cm]{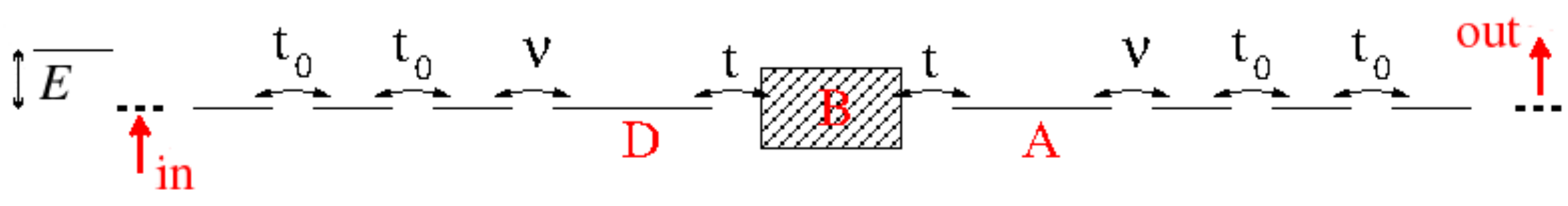}
\caption{The D-B-A system attached via a coupling $\nu$ to tight-binding
electrodes with electron hopping $t_0.$ An energy $E$ is imparted to the transferred
electron at the far-left. Shown are the locations, deep inside
the electrodes, where the incoming and outgoing quantities are probed.}
\label{wire}
\end{figure}

The D-B-A system coupled to electrodes was set up as in Fig~\ref{wire}. At time zero, The 
transferred electron is imparted with an energy $E$ on a site
at the far end of the left electrode while the wavepacket amplitude or current is calculated
at a reference point located deep inside
the electrode in order to avoid the effects of reflection from the bridge. The location of the transmission
probe had no effect. The
effect of the electrodes is minimal when the coupling to the D-B-A system (hopping $\nu$)
is small compared with the hopping $t_0$ in the wires.

The evolution of the wavefunction, using the TDSE, was performed using finite electrodes, each numbering 500 sites, longer
than necessary for the results to converge. While using lattice Green's functions, the length
of the electrodes can in principle be extended to infinity, the ranges must nonetheless be truncated in space if the 
interaction is to be maintained in its exact form. By
using finite electrodes it is a simple matter to impart an energy $E$ to the transferring electron at
the left-most site, as seen in Fig.~\ref{wire}. As it is localized on a non-eigenstate, it transfers
even with no external field present. During the time evolution, the total energy of the system is conserved as 
is that of each electronic wavefunction separately since the electrons are are non-interacting. 

\subsection{Scattering-based versus time-dependent energy spectrum}
We compare our time-dependent calculation with that of a Fourier-transformed energy spectrum (scattering method) in which
the transmission is computed in terms of the scattered electron energy. In contrast, for the exact-diagonalization
approach, the expectation value of the energy is referred to since the energy of the extra electron does not
correspond to an eigenvalue of the system. The
Fourier-transform calculation of the transmission additionally differs from the explicit inclusion of the transferred electron as in
the `simple' calculation of $T_{DA}$: they are both lowest-order perturbation calculations. Thus the comparison 
will have both of these differences. However, for long bridges it is expected
that the difference arising from the use of the expectation value versus the exact energy will be 
minimal.  We calculate both the wavepacket and the
current transmission, comparing the time-dependent results with scattering theory.

\subsubsection{Wavepacket Transmission}
\begin{figure}[htb]
\includegraphics[width=14cm]{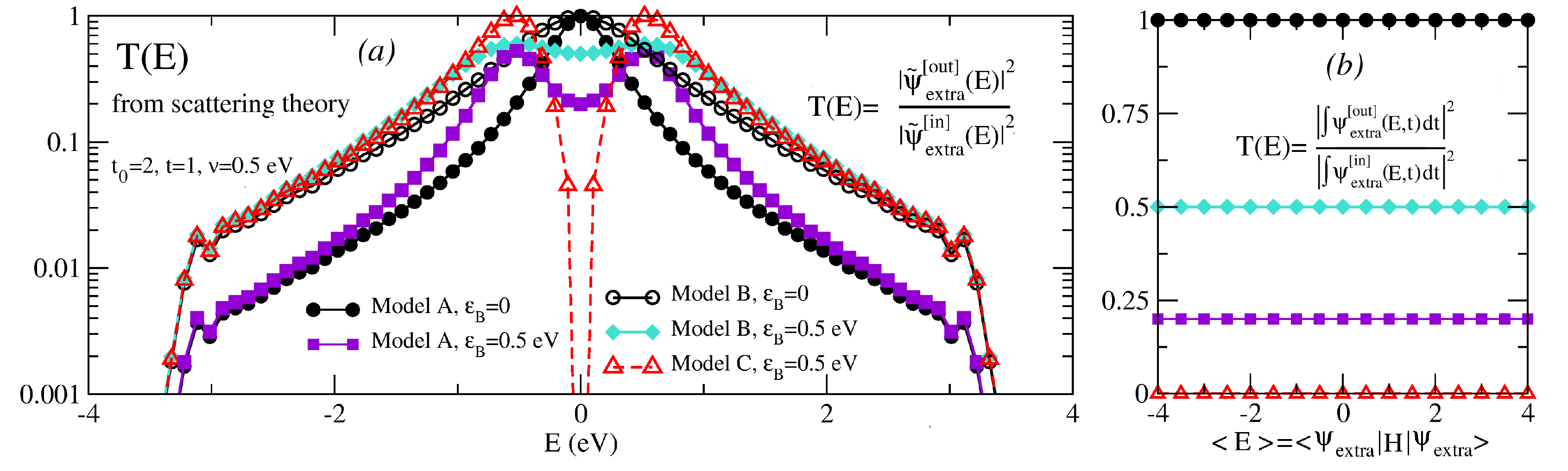}
\caption{Transmission of electron probability density across the bridge for various bridge-molecule configurations 
as indicated in the legend (see also Fig.~\ref{dba}) as a function
of the electron-scattering energy $E$ (a) or its expectation value $\langle E\rangle$ (b). Diagram (a): using
scattering theory (Fourier-Transform), indicated by the tilde (\textasciitilde) sign. Diagram (b): with exact diagonalization. The results
for Model A and Model B at $\epsilon_B=0$ coincide. The input/output values
were taken at 100 electrode sites before/after the bridge (schematic in Fig.~\ref{wire}).}
\label{FT}
\end{figure}

In Fig.~\ref{FT} we show the transmission of the probability density
for a variety of molecule configurations as a function
of scattering energy $E$. The following geometries were studied: a single molecule, and two molecules in the parallel-path
configuration oriented either with the same or opposite energies, as in Fig.~\ref{dba}. In the exact approach the energy-dependent
probability density is obtained by integrating the strongly-peaked wavepacket over a window: $\left|\int\psi_{extra}(E,t)\,\mathrm{dt}\right|^2$
while in the scattering approach it
is expressed as a Fourier-Transform: $\left|\tilde{\psi}_{extra}(E)\right|^2.$

In the scattering approach, the height and location of the peaks in the transmission in Fig.~\ref{FT}a are determined by the 
bridge-molecule characteristics: energy of the bridge
as well its coupling to the electrodes. The latter is kept small in order to minimize the
effect of the electrodes. The effect of two parallel bridges is to increase the transmission of the 
scattered electron at $E=0$ when the bridge-molecule energy is different from that of the wires; the larger
the bridge-molecule energy is the closer the transmission gets to achieving a doubling of its value
in the single bridge case. For oppositely-configured bridges the transmission is zero at $E=0$ (anti-resonance). However, 
at non-zero $E$, the transmission can be enhanced compared to the single-bridge case. Generally, the double bridges
lead to enhanced transmission over a wide range of scattering energies of the electron. The results
compare well with the existing literature~\cite{PickupFowler,Solomon,Renaud} while the band edge at $\pm 2t_0$ 
becomes a sharp cutoff in
the transmission as the length of the electrodes approaches infinity (not shown).

For the explicitly time-dependent calculations, the results are completely different, as can be seen in 
Fig.~\ref{FT}b, the two agreeing only at zero electron energy. Rather
than displaying resonances and anti-resonances, the spectrum is completely flat, as would
be expected from the frequency response to an impulse input. Rather than being dominated
by the structure of the D-B-A system, the frequency response is determined by the transferred electron,
with the D-B-A system imparting only a constant modification. The destructive interference
in model C in particular holds for all energies. The effect of parallel, equivalent bridges (model B) off resonance 
is to more than double the transmission compared to a single bridge (model A), an effect previously found in conductance
experiments~\cite{Vazquez,Kiguchi}. One study found that electronic effects (functional groups) do not greatly 
modify single-molecule conductance~\cite{Mowbray} while geometrical (conformation) effects do have some effect~\cite{NatureConformation}.

\subsubsection{Current Transmission}
\label{sec:CurrentDensity}
As the wavepacket transmission is not normally observed in experiments, we choose to study the
current transmission.
The continuous-current formula
\begin{equation}
\frac{\partial\rho}{\partial t}+\nabla\cdot\vec{j}=0
\end{equation}
has the discretized form at the link between the $n$th and
the $n+1$th sites (separated by the lattice spacing $a$) of a tight-binding chain~\cite{discrete}
\begin{equation}
j_{n,n+1}=\frac{i}{\hbar}at_0(\psi_{n+1}^*\psi_n-\psi_n^*\psi_{n+1})
\label{discr}
\end{equation}
whence the current at the energy $E$ is given by
\begin{equation}
J(E)=\int_{t=0}^{\infty} j(E,t)\,dt.
\label{transmission}
\end{equation}
The current is found to be very strongly-peaked and Eq.~\ref{transmission} is thus integrated over a window so as not
to be influenced by reflection from the wire ends. Although
some authors~\cite{Bishop} consider that Eq.~\ref{transmission} describes the transmission
$T(E)$ directly, we label it as $J(E)$ as in our case we calculate the transmission
with respect to a point deep inside the left electrode (Fig.~\ref{wire})
and we thus need to normalize it in terms of that quantity.

\begin{figure}[htb]
\includegraphics[width=14cm]{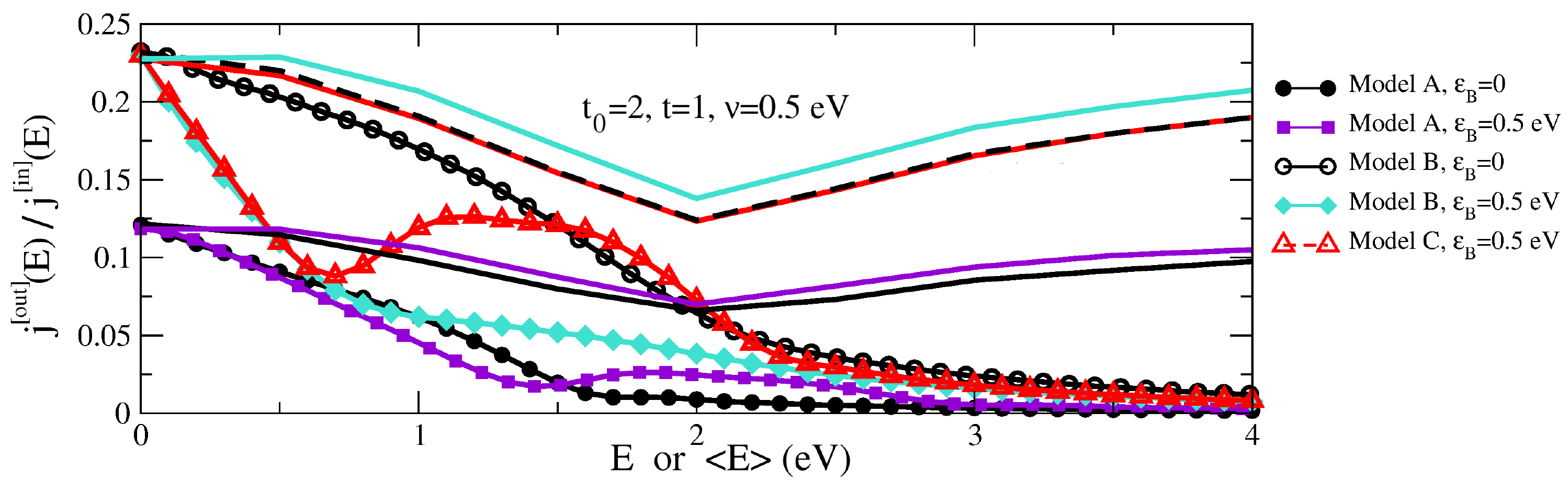}
\caption{Transmission of current across the bridge for various bridge-molecule configurations as a function
of energy $E$. The coupling parameter of the electrode to the bridge is $\nu=0.5.$ Compared
are the scattering (symbols) and the exact (bare lines) approaches. Otherwise the colour correspondence
shown in the legend applies to both approaches, only that the bare dashed black line (exact approach) 
corresponds to the case of Model B, $\epsilon_B=0$ (open circle in scattering approach).}
\label{jfig}
\end{figure}

We compare the current according to Eq.\ref{transmission} for the exact, time-dependent case (with
the expectation value of the energy) with the Fourier-transformed current, which is appropriate
for a steady-state situation, in Fig.~\ref{jfig}. As in the case of the wavepacket transmission, the results agree
only at zero energy. What is most striking in the comparison of the results obtained using the two 
methods is that while both begin by dropping off overall for increasing energy, in the
exact approach, the current transmission \emph{increases} at higher energies ($E>t_0$) and 
levels off. The reduction in the current transmission with increasing
$E$ (up to $t_0$, that is 2 eV) is due to two factors: off-resonant transmission in which the transmitted wavepacket
increasingly prefers to occupy lower density-of-states regions at the edges of the bands, which
are shaped as such on account of the defect D-B-A system, and secondly due to the back-reflection of incoming
and pinning of outgoing
electrons at the D-B-A system owing to the small coupling to electrodes $\nu$ compared with
the other hopping parameters. The latter is
the cause of the current transmission being small even at $E=0.$ Note that in this regard there is a substantive
difference with Fig.~\ref{FT}a which does reach values of 1. The quantity 
$\frac{\int \left|\psi_{\mathrm{extra}}^{\mathrm{[out]}}(E,t)\right|^2\,dt}{\int \left|\psi_{\mathrm{extra}}^{\mathrm{[in]}}(E,t)\right|^2\,dt},$
termed the \emph{occupation transmission}, resembles 
the `exact' curves in Fig.~\ref{jfig} qualitatively and quantitatively (not shown). 

At values of $E<t_0$ the current transmission
occurs via a `sequential' mechanism 
in which each site is distinctly occupied in sequence but at $E>t_0$ the
main mode of transmission is via a tunnelling mechanism via a defect state which splits off above
the conduction-band maximum which becomes increasingly favoured for higher $E.$ Because 
we are also in the direct ET regime in which the D-B-A system is now practically unoccupied, there 
is much less loss of transmission due to pinning or back-reflection, leading to the increase in current transmission. At 
$E=0$ the current transmission for the parallel bridge
configurations (models B and C) is double that of model A while there is no destructive
interference for model C. 

\section{Conclusions}
We have calculated the exact electron-transfer (ET) matrix elements for a few benchmark donor-bridge-acceptor systems
and we have found them to differ remarkably from the results obtained from perturbation theory. Rayleigh-Schr\"{o}dinger 
perturbation theory does not in general lead, at any order, to the exact 
results. Regardless, ET rates, which depend on the dominant Rabi frequency of the transfer
from donor to acceptor, are not representative of experimental measurements
of more relevant quantities such as current or wavepacket transmission / conductance. For our models, we find, in agreement
with some experiments, that current transmission is actually not very sensitive to the bridge
geometry but more to the number of parallel pathways. In the wavepacket transmission we do however find destructive
interference for one of the bridge configurations, but for all energies. In the current transmission we find constructive
but no destructive interferences resulting from the bridge-model geometries we studied.
For the wavepacket transmission we find a flat transmission spectrum in
the exact approach involving the explicit inclusion of the transferred electron as an impulse, in contrast
with the time-independent, scattering approach, which is appropriate to steady-state situations. 
The current transmission further displays an
unexpected characteristic for large energies: it increases as a result of tunnelling involving defect states
split off from the conduction-band maximum, an effect not captured by scattering theory where the existing
energy levels are not altered by the impinging electron.

\end{document}